# Fair Isaac Technical Paper

**Subject:** **Score Engineered Robust Least Squares Regression**

**From:** **Bruce Hoadley (BCH – A34 X27051)**

**Date:** **July 5, 2002**

## Abstract

In other FICO Technical Papers, I have shown how to fit Generalized Additive Models (GAM) with shape constraints using quadratic programming applied to B-Spline component functions. In this paper, I extend the method to Robust Least Squares Regression.

## Table of Contents

# 1. Introduction

The current version of INFORMedge provides the capability to develop score engineered weighted least squares regression models. These take the form of liquid scorecards or traditional scorecards. It is well known that the least squares method does not work well when there are outliers in the performance (dependent) variable. Outliers can exists in practical Fair, Isaac problems. For example, outliers are possible when the performance variable is revenue, loss, or change in revenue.

Outliers can come in various flavors. For example, the conditional distribution of $y$ given $x$ might be very skewed to the right – as in the revenue example. Or the conditional distribution of $y$ given $x$ might be symmetric, but have thick tails – as in the change in revenue example.

The different flavors of outliers require different statistical treatment. This paper covers the case where the conditional distribution of $y$ given $x$ is roughly symmetric, but has thick tails. In this case, the appropriate method is robust least squares regression.

Version 2 of the INFORMedge MATLAB code now has an implementation of robust least squares regression. This paper provides a mathematical documentation of the algorithms used in that implementation. The fitting algorithm is based on the Huber loss function, and is sometimes called M-Regression. The algorithm is an adaptation of the algorithm presented in Chapter 5 of Reference [9]. One of the adaptations was for the sake of score engineering.

I also document the associated marginal contribution algorithms. These are heuristics based on a winsorized version of the marginal contribution algorithms used in regular least squares.

A traditional Fair, Isaac approach, for handling outliers in a continuous outcome, is to transform to the binary outcome domain via truncation and parceling. A comparison of the new approach to the traditional approach would make for interesting research.

## 2. Score Engineered Least Squares Regression

I start this paper by reviewing the score engineered regression formulation introduced in Section 6 of Reference [1].

### 2.1. Mathematical formulation

In a regression problem there is a dependent variable, $y$, which usually takes on a variety of numerical values. Revenue is an example of a dependent variable. For linear regression, the model fit to the dependent variable is of the form

$$E[y \mid X] = S_0 + \sum_{j=1}^{p} S_j X_j ,$$

where the $X_j$'s are the independent variables and the $S_j$'s are the regression coefficients. In the case where the $X_j$'s are attribute indicator variables, the scorecard part of the final model is $\sum_{j=1}^{p} S_j X_j$ .

The score engineered regression problem can be formulated as

$$\begin{aligned}
&\text{Find } S_0 \text{ and } \boldsymbol{S} \text{ to} \\
&\text{Minimize } \sum_{i=1}^{n} w_i \left( y_i - \left( S_0 + \sum_{j=1}^{p} S_j x_{ij} \right) \right)^2 + \frac{\lambda}{n} \boldsymbol{S}' * \boldsymbol{S} \\
&\text{Subject to:} \\
&\quad \boldsymbol{Ai} * \boldsymbol{S} = \boldsymbol{IW} \\
&\quad \boldsymbol{Ac} * \boldsymbol{S} = 0 \\
&\quad \boldsymbol{Ap} * \boldsymbol{S} \le 0 ,
\end{aligned}$$

where

$$\sum_{i=1}^{n} w_i = 1 .$$

Note that this formulation is slightly different than the formulation presented in Section 6 of Reference [1]. First I have added the sample weights, $w_i$. Also, the penalty parameter, $\lambda$, (model.penalty in INFORMedge) is divided by *n* rather than *p*. The reason is explained in Reference [8]. With this formulation, the value of the penalty parameter can be specified independently of *n* and *p.*

The score engineered regression problem can be put into matrix notation by defining some new matrices. Let

$$\beta' = \begin{bmatrix} S_0 & \boldsymbol{S}' \end{bmatrix}$$

$$\boldsymbol{y}' = \begin{bmatrix} y_1, y_2, \ldots, y_n \end{bmatrix}$$

$$\boldsymbol{Xr} = \begin{bmatrix} 1 & x_{11} & x_{12} & \bullet & \bullet & x_{1p} \\ 1 & x_{21} & & & & x_{2p} \\ \bullet & \bullet & & x_{ij} & & \bullet \\ \bullet & \bullet & & & & \bullet \\ 1 & x_{n1} & x_{n2} & \bullet & \bullet & x_{np} \end{bmatrix}$$

$$\boldsymbol{W} = \begin{bmatrix} w_1 & 0 & \bullet & 0 \\ 0 & w_2 & \bullet & 0 \\ \bullet & \bullet & \bullet & \bullet \\ 0 & 0 & \bullet & w_n \end{bmatrix}$$

$$\boldsymbol{Ir} = \begin{bmatrix} 0 & 0 & 0 & 0 \\ 0 & 1 & 0 & 0 \\ 0 & 0 & 1 & 0 \\ 0 & 0 & 0 & 1 \end{bmatrix} \quad ((p+1) \times (p+1) \text{ matrix})$$

$$\boldsymbol{Air} = \begin{bmatrix} \boldsymbol{0} & \boldsymbol{Ai} \end{bmatrix}$$

$$\boldsymbol{Acr} = \begin{bmatrix} \boldsymbol{0} & \boldsymbol{Ac} \end{bmatrix}$$

$$\boldsymbol{Apr} = \begin{bmatrix} \boldsymbol{0} & \boldsymbol{Ap} \end{bmatrix}.$$

In this notation, the score engineered regression problem is

$$\begin{aligned}&\text{Find } \boldsymbol{\beta} \text{ to}\\&\text{Minimize } (\boldsymbol{y}-\boldsymbol{Xr}*\boldsymbol{\beta})'*\boldsymbol{W}*(\boldsymbol{y}-\boldsymbol{Xr}*\boldsymbol{\beta})+\frac{\lambda}{n}\boldsymbol{\beta}'*\boldsymbol{Ir}*\boldsymbol{\beta}\\&\text{Subject to:}\\&\quad \boldsymbol{Air}*\boldsymbol{\beta}=\boldsymbol{IW}\\&\quad \boldsymbol{Acr}*\boldsymbol{\beta}=\boldsymbol{0}\\&\quad \boldsymbol{Apr}*\boldsymbol{\beta}\le\boldsymbol{0}.\end{aligned}$$

The first part of the objective function can be expanded as follows:

$$\begin{aligned}&(\boldsymbol{y}-\boldsymbol{Xr}*\boldsymbol{\beta})'*\boldsymbol{W}*(\boldsymbol{y}-\boldsymbol{Xr}*\boldsymbol{\beta})\\&\quad=\boldsymbol{\beta}'*\boldsymbol{Xr}'*\boldsymbol{W}*\boldsymbol{Xr}*\boldsymbol{\beta}-2*\boldsymbol{y}'*\boldsymbol{W}*\boldsymbol{Xr}*\boldsymbol{\beta}+\boldsymbol{y}'*\boldsymbol{W}*\boldsymbol{y}.\end{aligned}$$

Since $\boldsymbol{y}'*\boldsymbol{W}*\boldsymbol{y}$ is a constant, it can be dropped from the objective function. The result is

**Score engineered regression quadratic program**

$$\begin{aligned}&\text{Find } \boldsymbol{\beta} \text{ to}\\&\text{Minimize } \boldsymbol{\beta}'*\left(\boldsymbol{Xr}'*\boldsymbol{W}*\boldsymbol{Xr}+\frac{\lambda}{n}\boldsymbol{Ir}\right)*\boldsymbol{\beta}-2*\boldsymbol{y}'*\boldsymbol{W}*\boldsymbol{Xr}*\boldsymbol{\beta}\\&\text{Subject to:}\\&\quad \boldsymbol{Air}*\boldsymbol{\beta}=\boldsymbol{IW}\\&\quad \boldsymbol{Acr}*\boldsymbol{\beta}=\boldsymbol{0}\\&\quad \boldsymbol{Apr}*\boldsymbol{\beta}\le\boldsymbol{0}.\end{aligned}$$

### 2.2. MATLAB formulation of the score engineered regression quadratic program

The general form of the MATLAB quadratic program is

$$\begin{aligned}&\text{Find } \boldsymbol{\beta} \text{ to}\\&\text{Minimize } \left(\frac{1}{2}\right)\boldsymbol{\beta}'*\boldsymbol{H}*\boldsymbol{\beta}+\boldsymbol{f}'\boldsymbol{\beta}\\&\text{Subject to:}\\&\quad \boldsymbol{Aeq}*\boldsymbol{\beta}=\boldsymbol{beq}\\&\quad \boldsymbol{A}*\boldsymbol{\beta}\le\boldsymbol{b}\\&\quad \boldsymbol{l}\le\boldsymbol{\beta}\le\boldsymbol{u}.\end{aligned}$$

To facilitate the description of the algorithm in Section 3, the solution to this quadratic program function is denoted in functional form as

$$\boldsymbol{\beta} = QP(\boldsymbol{H}, \boldsymbol{f}, \boldsymbol{A}, \boldsymbol{b}, \boldsymbol{Aeq}, \boldsymbol{beq}, \boldsymbol{l}, \boldsymbol{u}).$$

So, for the score engineered regression quadratic program, the matrices in the general form of the MATLAB quadratic program are

$$\boldsymbol{H} = 2\left(\boldsymbol{Xr}' * \boldsymbol{W} * \boldsymbol{Xr} + \frac{\lambda}{n}\boldsymbol{Ir}\right)$$
$$\boldsymbol{f} = -2(\boldsymbol{Xr}' * \boldsymbol{W} * \boldsymbol{y})$$
$$\boldsymbol{Aeq} = \begin{bmatrix} \boldsymbol{Air} \\ \boldsymbol{Acr} \end{bmatrix}$$
$$\boldsymbol{beq} = \begin{bmatrix} \boldsymbol{IW} \\ 0 \end{bmatrix}$$
$$\boldsymbol{A} = \boldsymbol{Apr}$$
$$\boldsymbol{b} = 0$$
$$\boldsymbol{l} = -\infty$$
$$\boldsymbol{u} = +\infty.$$

# 3. Robust Least Squares Regression

## 3.1 Huber loss function

In regular least squares regression one tries to minimize squared error loss. In the language of statistical decision theory, the per observation loss function is

$$L(e) = e^2,$$

where $e$ is a residual error.

This method does not work well when there are outliers in the $\boldsymbol{y}$ variable. A popular robust alternative to squared error loss is the Huber loss function

$$\rho(e) = \begin{cases} e^2 & \text{if } |e| \le k \\ 2ke - k^2 & \text{if } e > k \\ -2ke - k^2 & \text{if } e < k, \end{cases}$$

where $k$ is a multiple of a robust estimate of the residual standard deviation. The shape of $\rho(e)$ looks similar to the shape of $L(e)$, but there is a big difference. When $|e| > k$, the value of $\rho(e)$ increases linearly in $e$, whereas the value of $L(e)$ increases as a quadratic in $e$. Regression coefficients that minimize a weighted sum of the Huber losses is more robust than regression coefficients that minimize a weighted sum of squared errors.

### 3.2 Algorithm

The algorithm for score engineered robust least squares regression, based on the Huber loss function, is an iterative version of the score engineered least squares regression in Section 2. The basic concepts of the iterative algorithm are derived on p. 88 of Reference [9]. Of course, Reference [9] does not deal with score engineered regression, but the generalization is easy. For each iteration, a winsorized version of the dependent variable, $\boldsymbol{y}$, is computed, and then score engineered least squares regression is performed using this winsorized $\boldsymbol{y}$.

Before describing the algorithm, I introduce some notation. A line starting with the symbol, %, is a comment on the algorithm. The term, wtmedian $(\boldsymbol{x}, \boldsymbol{w})$, is the weighted median of the variable $\boldsymbol{x}$ using the weights $\boldsymbol{w}$. The symbol $\bullet *$ indicates element-by-element vector multiplication.

Here is a description of the score engineered robust least squares algorithm.

% Specify input parameters

$M$ = Maximum number of iterations allowed

$\varepsilon$ = Upper bound on maximum absolute change in regression coefficients. This is used to control convergence.

$m$ = Multiple of robust standard deviation used to define the winsorization of $\boldsymbol{y}$ ( the default value is 1.5).

% Compute initial $\boldsymbol{\beta}$ via score engineered least squares regression

$$\boldsymbol{\beta} = QP(\boldsymbol{H}, \boldsymbol{f}, \boldsymbol{A}, \boldsymbol{b}, \boldsymbol{Aeq}, \boldsymbol{beq}, \boldsymbol{l}, \boldsymbol{u})$$

% Set up the while loop

$$c = 1$$

$$\boldsymbol{\beta_o} = \boldsymbol{\beta}$$

$$\beta_o(1) = \beta(1) + 2\varepsilon$$

% Execute the while loop

while $(\max(|\boldsymbol{\beta_o} - \boldsymbol{\beta}|) > \varepsilon) \,\&\, (c < M)$

$$\boldsymbol{\beta_o} = \boldsymbol{\beta}$$

$$c = c + 1$$

$$\boldsymbol{ae} = |\boldsymbol{y} - \boldsymbol{Xr} * \boldsymbol{\beta}|$$

$$\sigma = (1.483) * \text{wtmedian}(\boldsymbol{ae}, \boldsymbol{w})$$

$$k = m * \sigma$$

$$e_i^* = \begin{cases} e_i & \text{if } |e_i| \le k \\ k & \text{if } e_i > k \\ -k & \text{if } e_i < -k \end{cases}$$

$$\boldsymbol{y}^* = \boldsymbol{Xr} * \boldsymbol{\beta} + \boldsymbol{e}^*$$

$$\boldsymbol{f} = -2 * \boldsymbol{Xr}' * (\boldsymbol{w} \bullet * \boldsymbol{y}^*)$$

$$\boldsymbol{\beta} = QP(\boldsymbol{H}, \boldsymbol{f}, \boldsymbol{A}, \boldsymbol{b}, \boldsymbol{Aeq}, \boldsymbol{beq}, \boldsymbol{l}, \boldsymbol{u})$$

end

Note that $\sigma$ is a traditional robust estimate of the residual standard deviation. When the residuals are normally distributed, this estimate is unbiased.

It is shown in Chapter 5 of Reference [9] that this algorithm minimizes the Huber loss function.

## 4. Marginal Contributions

A direct use of the Huber objective function (loss function) does not work for the marginal contribution calculations. Believe me, I tried it and failed. The reason is that outliers in ***y*** can dominate the average Huber loss function values – even though the Huber loss increases linearly in large errors. To get around this problem, I have developed a heuristic for the marginal contribution calculations in the robust least squares case. The heuristic is based on applying the regular least squares approach to the winsorized ***y*** that comes out of the Step I fitting process.

The algorithms below can be applied to the development or validation samples. This feature is available in Version 2 of the INFORMedge MATLAB code.

**4.1 Step I**

The last iteration of the Step I fitting process produces winsorized errors and outcomes, $\boldsymbol{e}^*$ and $\boldsymbol{y}^*$. From this, I can compute a winsorized sum of squared errors

$$SSE^* = \sum_{i=1}^{n} w_i \left(e_i^*\right)^2 .$$

To normalize this, I use a winsorized variance of y

$$RLSVy = \sum_{i=1}^{n} w_i \left(y_i^* - \beta(1)\right)^2 .$$

Note that $\beta(1)$ acts like a robust estimate of the mean of ***y***, because in INFORMedge, the characteristic scores are all centered on zero. So the intercept term is a robust estimate of the mean of ***y.***

The heuristic objective function value associated with the fitted Step I score is

$$OF = \frac{SSE^*}{RLSVy} .$$

Now we want to compute the Step I marginal contribution for a particular Step I characteristic. First we set the score weights, associated with the Step I characteristic, equal to zero to yield $\boldsymbol{\beta I}$. Next we compute the associated winsorized error vector

$$\boldsymbol{eI} = \boldsymbol{y}^* - \boldsymbol{Xr} * \boldsymbol{\beta I}.$$

Next we compute the winsorized-weighted sum of squared errors

$$SSEI = \sum_{i=1}^{n} w_i (eI_i)^2 .$$

Finally, the Step I marginal contribution is

$$MCI = \frac{SSEI}{RLSVy} - OF.$$

**4.2 Step II**

Again, a direct use of the Huber objective function (loss function) does not work for the marginal contribution calculations. So again I use a heuristic.

The last iteration of the Step I fitting process produces winsorized outcomes, $\boldsymbol{y}^*$.

Consider a particular Step II characteristic. Our approach is to fit a Step II robust model, where the independent variables are the spline basis functions for the Step II characteristic – supplemented by the score variable, ***s***, that came out of the Step I fit. To get robustness, we just apply least squares using the winsorized $\boldsymbol{y}^*$. For the Step II model, we use an intercept term, and we constrain the regression coefficient of ***s*** to be 1. We then use this model to score out the sample to yield the Step II score variable, ***sII***.

Next we compute the associated winsorized error vector

$$\boldsymbol{eII} = \boldsymbol{y}^* - \boldsymbol{sII}.$$

Next we compute the winsorized-weighted sum of squared errors

$$SSEII = \sum_{i=1}^{n} w_i (eII_i)^2 .$$

Finally, the Step II marginal contribution is

$$MCII = OF - \frac{SSEII}{RLSVy}.$$